\documentclass[aps,twocolumn,showpacs,amsmath,final,amssymb]{revtex4-1}
\usepackage{float}
\usepackage{inputenc}
\usepackage{amsmath}
\usepackage[dvips,final]{graphicx}
\usepackage{color}
\renewcommand{\arraystretch}{1.2}

\usepackage{enumitem}
\usepackage{color}

\begin{document}

\title{Exchange coupling and magnetic anisotropy at Fe/FePt interfaces}

\author{C J Aas,$^1$ P J Hasnip,$^{1}$ R Cuadrado,$^{1}$ E M Plotnikova,$^{1}$ L Szunyogh,$^{2,3}$ L Udvardi,$^{2,3}$  and R W Chantrell$^1$}
\affiliation{$^1$ Department of Physics, University of York, York YO10 5DD, United Kingdom}
\affiliation{$^2$ Department of Theoretical Physics, Budapest University of Technology and Economics, Budafoki \'ut 8.~H1111 Budapest, Hungary}
\affiliation{$^3$ Condensed Matter Research Group of Hungarian Academy of Sciences, Budapest University of Technology and Economics, Budafoki \'ut 8.,~H-1111 Budapest, Hungary }

\date{\today}

\begin{abstract}
We perform fully relativistic first principles calculations of the exchange interactions and the magnetocrystalline anisotropy energy (MAE) in an Fe/FePt/Fe sandwich system in order to elucidate how the presence of Fe/FePt (soft/hard magnetic) interfaces impacts on the magnetic properties of Fe/FePt/Fe multilayers. Throughout our study we make comparisons between a geometrically unrelaxed system and a geometrically relaxed system.  We observe that the Fe layer at the Fe/FePt interface plays a crucial role inasmuch its (isotropic) exchange coupling to the soft (Fe) phase of the system is substantially reduced.  Moreover, this interfacial Fe layer has a substantial impact on the MAE of the system.  We show that the MAE of the FePt slab, including the contribution from the Fe/FePt interface, is dominated by anisotropic inter-site exchange interactions.   Our calculations indicate that the change in the MAE of the FePt slab with respect to the corresponding bulk value is negative, i.e., the presence of Fe/FePt interfaces appears to reduce the perpendicular MAE of the Fe/FePt/Fe system.  However, for the relaxed system, this reduction is marginal.  It is also shown that the relaxed system exhibits a reduced interfacial exchange. Using a simple linear chain model we demonstrate that the reduced exchange leads to a discontinuity in the magnetisation structure at the interface.
\end{abstract}


\maketitle

\section{Introduction}
\label{sec:intro}
Exchange-coupled soft/hard composite magnetic systems are of significant interest for their potential application in many different fields of technology such as magnetic recording media \cite{storage}, permanent magnets \cite{permanent} and magnetic microactuators \cite{actuators}.   A wealth of different soft/hard materials has been investigated in the literature (see, e.g., \cite{hardsoft1,hardsoft2,hardsoft3,hardsoft4,hardsoft5,hardsoft6,hardsoft7}).   Experimentally, the Fe/FePt system is a highly suitable system for studying the fundamental properties of nano-composite magnetic systems as the properties are relatively easy to control \cite{composites}.  Moreover, due to the high magnetisation of the saturated $\alpha$-Fe phase and the large magnetocrystalline anisotropy energy (MAE) of the FePt $L$1$_0$ phase, Fe/FePt bilayers are considered an ideal structure for exchange spring behaviour \cite{hawig} and for application in exchange-coupled (ECC) magnetic recording media.  For ECC applications, the (soft) Fe phase, through its exchange interaction with the (hard) FePt phase, would act as a ``lever'', reducing the write field.  Meanwhile, the thermal stability of the written information would be ensured by the large MAE of $L$1$_0$ FePt.  Thus, in order to realise such devices, the MAE of FePt needs to be maintained (if possible, enhanced). The effect of the Fe/FePt interface on the FePt MAE is, therefore, a very important aspect.

The aim of the present work is to investigate in detail the effect of the Fe/FePt interface on the exchange coupling and the MAE of an Fe/FePt/Fe system by means of first-principles calculations.  We compare the results of a geometrically relaxed Fe/FePt/Fe system to the corresponding results for an unrelaxed such system.  The latter is similar to the system studied by Sabiryanov and Jaswal \cite{sabiryanov}.  We use CASTEP \cite{CASTEP1, CASTEP2, CASTEP3} to obtain the relaxed ionic coordinates of an Fe/FePt/Fe system.  We then employ the fully relativistic screened Korringa-Kohn-Rostoker (SKKR) method \cite{KKR1} to calculate tensorial exchange interactions and the layer-resolved contributions to the MAE of the relaxed and unrelaxed Fe/FePt/Fe structures. Moreover, we evaluate the change in the FePt MAE induced by the presence of the Fe/FePt interfaces.

Such a study is not only  important from the point of view of understanding the properties of this nano-composite, but the site-resolved information is also central to the development and parameterisation of localised-spin models.  This strategy has recently been realized to study an IrMn$_3$/Co(111) interface.~\cite{IrMnCo} The exchange interactions and the on-site magnetic anisotropy constants have been calculated in the antiferromagnetic and ferromagnetic parts of the system, as well as at the interface between them, and then used in atomistic spin-dynamics simulations to investigate the exchange bias effect.~\cite{IrMnCo-EB} In particular, it was found that the exchange bias effect in this system is mainly  governed by large Dzyaloshinskii-Moriya (DM) interactions~\cite{dzyalo,moriya} between the Mn and Co atoms at the interface.

We thus consider the implications of our calculated ab-initio parameters in a multiscale modelling approach. Here one maps ab-initio information onto a fixed-spin atomistic model to allow calculations of thermodynamic quantities and magnetization dynamics.  Using a simple mapping onto a linear chain we study domain structures at the FePt/Fe interface using the ab-initio parameters. Importantly, the ab-initio calculations show that the relaxed system exhibits a reduced interfacial exchange coupling. Mapping this information onto the linear chain model shows that this reduction gives rise to a discontinuity in the magnetization at the FePt/Fe interface. The implications for the exchange spring effect are considered. The functionality of magnetic materials increasingly relies on structural design at the nanoscale, the exchange spring phenomenon, its use in permanent magnets and recording media being an excellent example. Mesoscopic calculations often assume bulk exchange coupling across interfaces, which clearly may not be the case, and will certainly depend on the material properties in a way which can only be elucidated by electronic structure calculations.

\section{Details of the calculations}\label{sec:calcs}
First we describe the geometric structure of the Fe/FePt/Fe sandwich system that we have chosen
for our investigation.
The SKKR method requires the system to be considered in terms of an \emph{interlayer} region (region $I$) positioned between two semi-infinite bulk regions. For region $I$, we considered the following sequence of atomic layers (AL) as shown in Table \ref{table:IL-bulkFe}: 7 Fe AL + 17 Pt/Fe/$\cdots$/Fe/Pt AL + 7 Fe AL, enclosed by two semi-infinite bulk Fe systems.\\*

 \begin{table*}[htp!]\centering
\begin{tabular}{@{}cccccccccccccccccc@{}}\hline
\vspace{0.2cm}
  $\cdots$  & -10 & -9 &  -8 &  -7 &  -6 &   $\cdots$ &  -1 &  0 &  1 &  $\cdots$ &  6 &  7 &  8 &  9 &  10 &  $\cdots$  \\ \hline
\vspace{0.2cm}
 $\cdots$ & \underline{ Fe} &   \underline{ Fe} &  \textbf{ Pt} &  \textbf{ Fe} &  \textbf{ Pt} &   $\cdots$ & \textbf{ Fe} & \textbf{ Pt} & \textbf{ Fe} &  $\cdots$ & \textbf{ Pt} &  \textbf{ Fe} &  \textbf{ Pt} &  \underline{ Fe} &  \underline{ Fe} &  $\cdots$  \\   \hline
\vspace{0.1cm}
\end{tabular}
\caption{The layout of the Fe/FePt/Fe structure enclosed by Fe bulk. Underlined chemical symbols refer to the Fe (soft magnet) part and bold face chemical symbols refer to the FePt (hard magnet) part of the system.\label{table:IL-bulkFe}}
\end{table*}

Using the layer sequence in Table \ref{table:IL-bulkFe}, we investigated the following two systems:

\begin{enumerate}[label=\textbf{\Alph*}]
\item A geometrically unrelaxed system with an overall two-dimensional lattice parameter, $a_{\mathrm{2D}} = a_{\mathrm{FePt}} / \sqrt{2} \approx 2.723$~\AA, where $a_{\mathrm{FePt}}=3.852$~\AA \ is the experimental in-plane lattice parameter of the $L$1$_0$ lattice of FePt.  Note that $2.723$ \AA~is within 5 \% of the experimental lattice parameter of bcc Fe,  $a^{(\mathrm{exp})}_{\mathrm{Fe}} =2.87$ \AA. For the FePt part of the system we used  the experimentally measured ratio of $c_{\mathrm{FePt}}/a_{\mathrm{FePt}}=0.964$,  while for the bcc Fe part, $c_{\mathrm{Fe}}=a_{\mathrm{2D}}$. At the interface between the Fe and FePt parts of the system, i.e., between AL's  $-9$ and $-10$, as well as between AL's 9 and 10 (see Table \ref{table:IL-bulkFe}), we set the interlayer separation to  $$c_{\mathrm{interface}} = \frac{c_{\mathrm{Fe}} + c_{\mathrm{FePt}}}{4} \; .$$
\item  The interlayer separations were relaxed using CASTEP, keeping $a_{\mathrm{2D}}=a_{\mathrm{Fe}}^{(\mathrm{LDA})}$, where $a_{\mathrm{Fe}}^{(\mathrm{LDA})}$ is the bulk lattice parameter for Fe obtained using Local Density Approximation (LDA) in CASTEP, $2.659$ \AA.  The resulting relaxed structure was then isotropically scaled up to the experimental FePt lattice parameter $a_{\mathrm{2D}}=2.723$ \AA \ in order to enable a direct comparison with the results for system {\bf A}. It should be noted here that the CASTEP geometry relaxation yields an Fe region which is slightly tetragonal (rather than cubic), with the ratio $c_{\mathrm{Fe}}/a_{\mathrm{Fe}} \approx 1.06$.   As this tetragonalisation must be due to the presence of the FePt slab, the relaxed geometry of system \textbf{B} corresponds more closely to a repeated multilayer structure.
\end{enumerate}

We should mention that we also investigated an unrelaxed sandwich system with a layout of $(1 \times 2)$ Fe/Pt AL + 9 Fe AL +  17 Pt/Fe/$\cdots$/Fe/Pt AL + 9 Fe AL + $(1 \times 2)$ Pt/Fe AL enclosed by FePt bulk (system {\bf C}). Since the considered Fe and FePt layers are quite thick, as expected, the magnetic properties (spin moments, exchange interactions and MAE) of system {\bf C} turned out very similar to those of system {\bf A}. In this study, we used system {\bf C} only for calibrating the change in the MAE of the FePt slab with respect to the MAE of bulk FePt.\\*

\begin{figure}[htp!]
\centering
\includegraphics[scale=0.56,trim=45mm 40mm 20mm 45mm, clip]{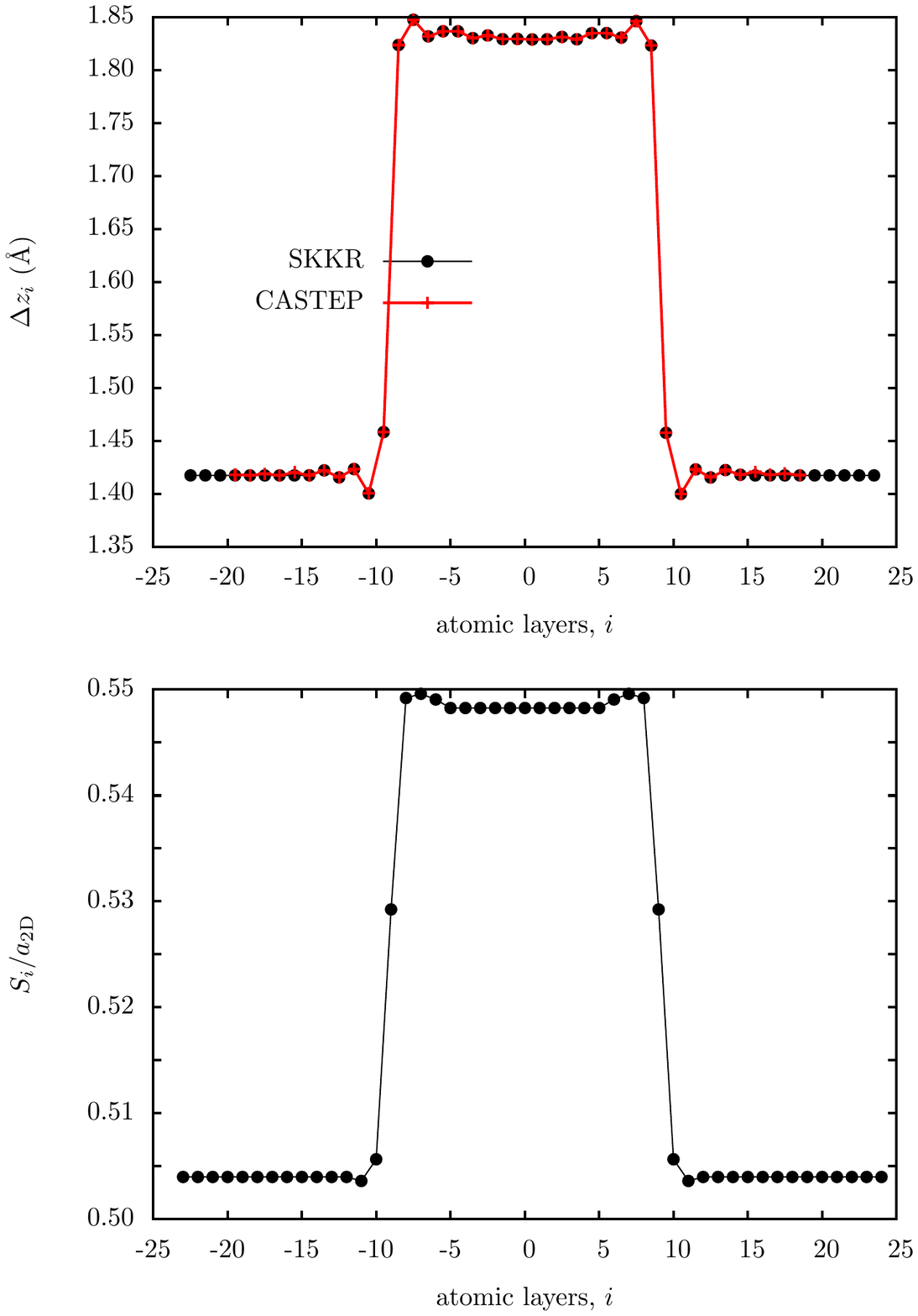}
% \scalebox{0.65}{\input{CASTEPFeFePtdeltaZ}}
% \scalebox{0.65}{\input{CASTEPFeFePtMT}}
\caption{Top: Interlayer spacings, $\Delta z_i$, and bottom: radii of atomic spheres, $S_i$ as used for system {\bf B}.
Red $+$ represent the interlayer spacings obtained from CASTEP, while black $\bullet$ represent the interlayer separations and the radii of atomic spheres as used in the SKKR calculations. Note that for the SKKR calculations the corresponding quantities are constant for layers $\left| i\right| > 16$. Solid lines serve as guides for the eye.
 \label{fig:CASTEPrel}}
\end{figure}

For system \textbf{B}, after specifying the vertical coordinate $z_i$ of each atomic layer $i$ from the CASTEP geometry relaxation, we needed to determine the corresponding atomic volumes, $V_i$.  Here the only strict requirement is that the sum of atomic volumes within region $I$ should be equal to the total lattice volume of region $I$, while the choice of the individual atomic volumes is somewhat arbitrary.  As a simple choice, we related the atomic volume of each atom in layer $i$, $V_i$, to the layer positions $\{z_i\}$ as $V_i=a_{\mathrm{2D}}^2 (z_{i+1}-z_{i-1})/2$. For Fe or FePt bulk, this construction trivially retains the corresponding bulk atomic volumes. For the case of system {\bf B}, in Fig.~\ref{fig:CASTEPrel} the interlayer distances, $\Delta z_i = z_{i+1} - z_i$, and the radii of the atomic spheres, $S_i$ (defined through $V_i=\frac{4\pi}{3}S^3_i$), are depicted according to the above construction. \\*

For each of the systems {\bf A} and {\bf B}, we performed self-consistent calculations by means of the SKKR method. We used the Local Spin-Density Approximation (LSDA) of the Density Functional Theory (DFT) as parameterized by Vosko {\it et al.} \cite{voskoCJP80}, with effective potentials and fields treated within the atomic sphere approximation (ASA).  The self-consistent calculations were performed within the scalar-relativistic approximation and an angular momentum cut-off  $\ell_{\mathrm{max}}=3$.\\*

The magnetocrystalline anisotropy energy (MAE) was then evaluated in terms of the fully relativistic SKKR method within the magnetic force theorem \cite{Jansen99}, in which the total energy of the system can be replaced by the single-particle (band) energy. Moreover, we employed the torque method \cite{MAEtorque1}, making use of the fact that, for a uniaxial system, the MAE, $K$, can be calculated up to second order in spin-orbit coupling as
\begin{equation}
 K = E(\theta=90^{\circ}) - E(\theta=0^{\circ}) = \left.\frac{dE}{d\theta}\right|_{\theta=45^{\circ}} \, ,
\label{eq:torque}
\end{equation}
where $\theta$ denotes the angle of the spin-polarization with respect to the $\left[0 0 1\right]$ direction of the FePt lattice. Within the KKR formalism, $K$ can be decomposed into layer-resolved contributions, $K_i$,
\begin{equation}
K = \sum_i  K_i \; .
\label{eq:Kdecomp}
\end{equation}
For more details on the torque method within the KKR method see Ref.~\cite{MAEtorque2}. We note that due to the two-dimensional translational symmetry of the systems, the MAE should be related to a 2D unit cell, therefore, in the following the index $i$ in Eq.~(\ref{eq:Kdecomp}) is used to label atomic layers. \\*

Having evaluated the layer-resolved contributions to the MAE for each Fe/FePt/Fe system, we considered next the potential mapping of such contributions to a localised-spin model.  Supposing that the electronic energy of a uniaxial magnetic system can be mapped into a generalised Heisenberg model,
\begin{equation}
\label{eq:Espinmodel}
{\cal H}=-\frac{1}{2}\sum_{i \ne j}\vec{S}_{i}\boldsymbol{J}_{ij}\vec{S}_{j}
-\sum_{i} d_i \left( \vec{S}_{i} \cdot \vec{e} \right)^2 \; ,
\end{equation}
where $\vec{S}_{i}$ represents a classical spin, i.e., a unit vector along the direction of the magnetic moment at site $i$. The first term stands for the exchange contribution to the energy, with $\boldsymbol{J}_{ij}$ denoting the tensorial exchange interaction, and the second term denotes the on-site anisotropy, with the anisotropy constant $d_i$ and the easy magnetic direction $\vec{e}$.  The exchange interaction matrix $\boldsymbol{J}_{ij}$  can further be decomposed into three terms \cite{spinwaves},
\begin{equation}\label{eq:Jij}
\boldsymbol{J}_{ij}=J_{ij}\boldsymbol{I}+\boldsymbol{J}_{ij}^{S}+\boldsymbol{J}_{ij}^{A} \; ,
\end{equation}
with $J_{ij}=\frac{1}{3} \mathrm{Tr}\boldsymbol{J}_{ij}$ the isotropic exchange interaction,  $\boldsymbol{J}_{ij}^{S}=\frac{1}{2}(\boldsymbol{J}_{ij}+\boldsymbol{J}_{ij}^{T})-J_{ij}^{iso}\boldsymbol{I}$ the traceless symmetric part of the exchange tensor and $\boldsymbol{J}_{ij}^{A}=\frac{1}{2}(\boldsymbol{J}_{ij}-\boldsymbol{J}_{ij}^{T})$ the antisymmetric part of the exchange tensor. \\*

Within the spin model, Eq.~(\ref{eq:Espinmodel}), the MAE of a uniaxial ferromagnetic system can be cast into on-site and inter-site parts,
\begin{equation}\label{eq:MAEspinmodel}
K = K_{\mathrm{on-site}} + K_{\mathrm{inter-site}} \; ,
\end{equation}
where
\begin{equation}\label{eq:MAEonsite}
K_{\mathrm{on-site}} = \sum_i d_i \; ,
\end{equation}
and
\begin{equation}\label{eq:MAEintersite}
K_{\mathrm{inter-site}} =  -\frac{1}{2}\sum_{i \ne j} \left( J^{xx}_{ij} - J^{zz}_{ij} \right) \; .
\end{equation}
Defining, thus, the layer-resolved inter-site anisotropy as
\begin{equation}
\label{eq:layerresKi}
K_{i,\mathrm{inter-site}} = -\frac{1}{2}\sum_{j (\ne i)} \left( J^{xx}_{ij} - J^{zz}_{ij} \right) \; ,
\end{equation}
the layer-resolved MAE in Eq.~(\ref{eq:Kdecomp}) can be compared within the spin model to
\begin{equation}\label{eq:Ki-spinmodel}
K_i = d_i + K_{i,\mathrm{inter-site}} \; .
\end{equation}
The exchange interaction matrices are calculated using the relativistic torque method as described in Ref.~\cite{spinwaves}.
The sum in Eq.~(\ref{eq:Ki-spinmodel}) over $j$ can be cast into sums over atomic layers and over sites within atomic layers. In particular, the latter one suffers from convergence problems since $\boldsymbol{J}_{ij}$ decays, at best, as $1/R_{ij}^3$, where $R_{ij}$  denotes the distance between atoms $i$ and $j$. For this reason the corresponding sum was transformed into an integral in $k$-space, the convergence of which could easily be controlled, for details see Ref.~\cite{spinwaves}.

\section{Results}
\subsection{Local spin moments}
\begin{figure}[htb!]
\centering
\includegraphics[scale=0.56,trim=45mm 40mm 20mm 45mm, clip]{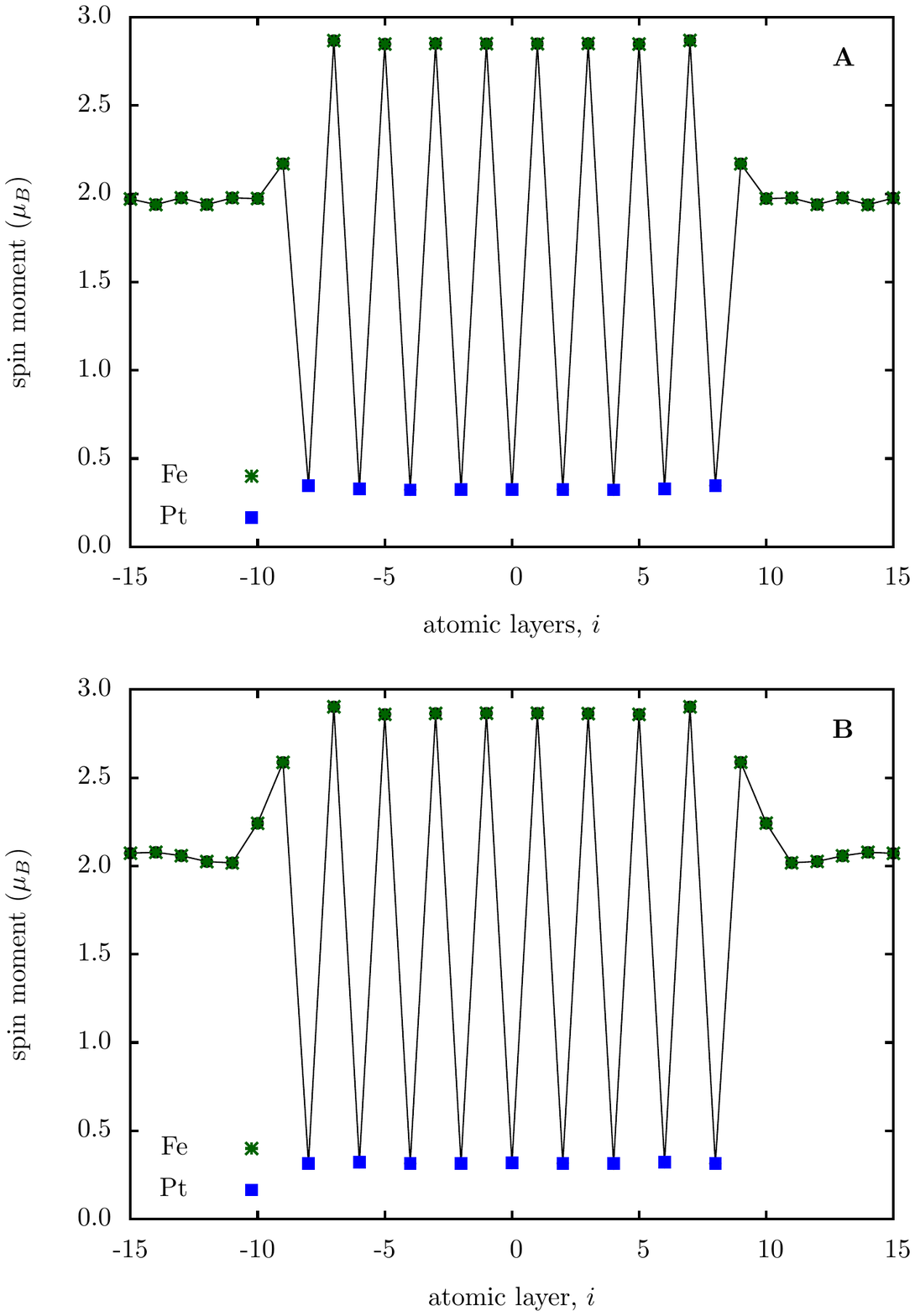}
% \scalebox{0.65}{\input{spinmom_UR}}
% \scalebox{0.65}{\input{spinmom_CASTEPlargea}}
\caption{Calculated layer-resolved spin moments (green $\ast$: Fe, blue $\blacksquare$: Pt) for systems {\bf A} and {\bf B}. Solid lines serve as guides for the eye.  \label{fig:spinmom}}
\end{figure}

The calculated atomic spin moments are plotted in Fig.~\ref{fig:spinmom}, displaying a fairly similar picture for the two systems {\bf A} and {\bf B}. In the interior of the FePt slab the moments,  $m_{\mathrm{Fe}}^{(\mathrm{FePt})}=2.86 \mu_{\mathrm{B}}$ and $m_{\mathrm{Pt}}^{(\mathrm{FePt})}=0.32\mu_{\mathrm{B}}$, are close to their bulk values in FePt.  Moreover, the Fe moments approach their bulk Fe value at the edges of region $I$.  Nevertheless, we observe that the bulk Fe spin moment is slightly enhanced in system {\bf B}, $m_{\mathrm{Fe}}^{(\mathrm{Fe, {\bf B}})}=2.07 \mu_{\mathrm{B}}$, as compared to system {\bf A}, $m_{\mathrm{Fe}}^{(\mathrm{Fe, {\bf A}})}=1.97 \mu_{\mathrm{B}}$.  This difference is, most likely, due to the slight tetragonality of the Fe unit cell along the $z$ direction (and the associated increase in volume) as discussed above for system {\bf B}.  An apparent difference between the spin moments of systems {\bf A} and {\bf B} occurs at the interface: although the Fe moments at the interface, $m_{\pm 9}$, are slightly increased in system {\bf A}, for system {\bf B} the enhancement of these moments is more pronounced.  Moreover, unlike in system {\bf A}, in system {\bf B} the spin moments in layer 10, $m_{\pm 10}$, are also enhanced. This means that, as expected, the transition of the moments from bulk Fe to bulk FePt is smoother in the relaxed case.\\*

\subsection{Effective exchange parameters}
\begin{figure}[ht]
\begin{center}
\includegraphics[scale=0.56,trim=45mm 40mm 20mm 45mm, clip]{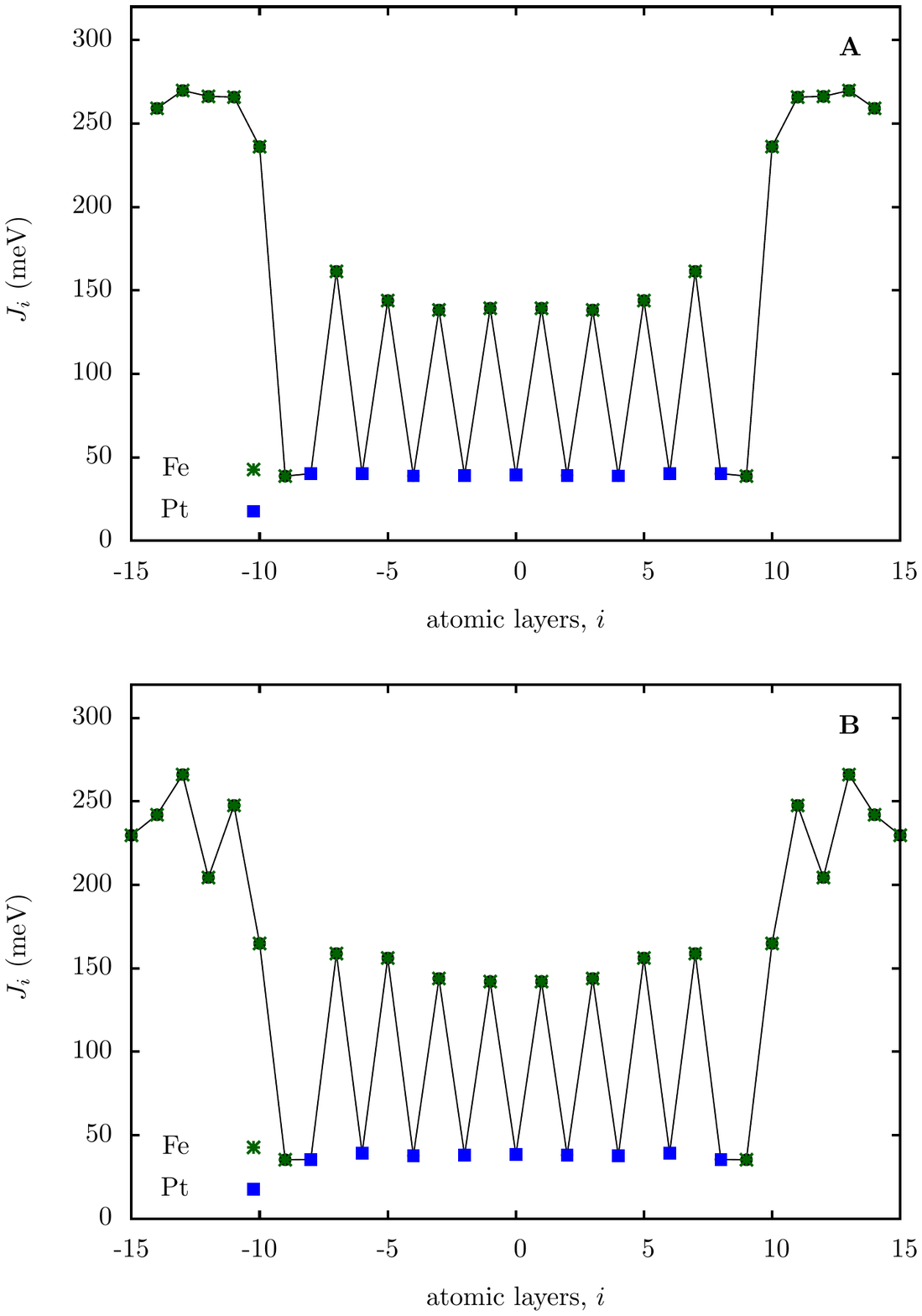}
% \scalebox{0.65}{\input{isotrexch_UR}}
% \scalebox{0.65}{\input{isotrexch_largeA}}
\end{center}
\vskip -8pt
\caption{Calculated layer-resolved effective isotropic exchange constants, $J_i$, see Eq.~(\ref{eq:Ji}). The blue $\blacksquare$ and green $\ast$ symbols represent $J_i$ for Pt and Fe layers, respectively. Solid lines serve as guides for the eye.}
\label{fig:isotrexchUR}
\end{figure}

In order to characterise the strength of the isotropic exchange interactions in a magnetic system, one often defines a site-resolved effective exchange parameter, $J_i$, defined for a given site $i$ as
\begin{equation} \label{eq:Ji}
 J_i = \sum_{j (\ne i)} J_{ij} \; .
\end{equation}
For a 2D translationally invariant system, $J_i$ must of course be identical for each site in a given layer.  Therefore, in the following, $i$ denotes the layer index.  For systems {\bf A} and {\bf B}, we calculated $J_i$ by considering all neighbours within a distance of seven $a_{\mathrm{2D}}$, which ensured a reliable convergence of the sum in Eq.~(\ref{eq:Ji}). The calculated layer-resolved effective exchange parameters are plotted in Fig.~\ref{fig:isotrexchUR} for systems {\bf A} and {\bf B}. \\*

Within the FePt slab, the effective exchange interactions of systems {\bf A} and {\bf B} exhibit very similar layer-resolved behaviours. The value of $J_{i}$ for the Fe layers is about 150~meV in the centre of the FePt slab and is slightly enhanced at the edges of the slab (i.e., towards layers $i = \pm 7$).  This is mainly a consequence of an enhancement of the ferromagnetic, nearest-neighbour (NN) intra-layer Fe-Fe interaction towards the outer layers of the FePt slab. The effective exchange parameter of about 40~meV observed in the Pt layers stems mainly from the strongly ferromagnetic nearest-neighbour Fe-Pt interactions. \\*

The (soft) Fe part of the system is characterised by much larger effective exchange, $J_i \sim 260$~meV.  This high value of the effective exchange parameter is a highly important property of the soft magnet part in exchange-coupled magnetic recording media as it enables the ``lever'' effect in switching the magnetisation of the hard (FePt) phase.  It should be noted that the effective exchange parameters calculated for the interior of the FePt slab and the Fe bulk part of the system correspond to mean-field Curie temperatures of $T^{\mathrm{MF}}_{\mathrm{C}} \sim 700 K$ and  $T_{\mathrm{C}}^{\mathrm{Fe}} \sim 1000 K$, in good agreement with the corresponding experimental values \cite{barmak,FeCurie}. \\*

Approaching the Fe/FePt interface from region $I$, the effective exchange of the Fe layers drops rapidly and the interfacial Fe layers $i = \pm 9$ exhibit an effective exchange of merely $\sim 40$ meV, almost identical to the effective exchange of the Pt layers. This reduction in $J_i$ originates in the weak interlayer couplings in layers $i = \pm 9$ and the relatively weak exchange of this layer with the soft layers $\left| i \right| \geq 10$.  In other words, what remains is essentially the ferromagnetic NN Fe-Pt interaction, thus giving Fe layers $i=\pm 9$ approximately the same effective exchange as the Pt layers.  Our results for the effective exchange parameters in the unrelaxed system {\bf A} are in satisfactory agreement with those in \cite{sabiryanov}, although the magnitudes of $J_i$ are significantly smaller in \cite{sabiryanov}. \\*

Interestingly, in system {\bf A}, the effective exchange of the Fe layer $i=\pm 10$, $J_{\pm 10} \sim 230$~meV, i.e., it almost recovers the bulk value.   In contrast, in system {\bf B}, $J_{\pm 10}$ remains remarkably small ($\sim 160$~meV).  Also, in system {\bf B} the effective exchange exhibits relatively large fluctuations throughout the Fe layers $\left| i \right| \geq 10$.  These differences could be attributed to the fact that the geometry of the Fe bulk is different for the two systems (see Section~\ref{sec:calcs}).  Although the oscillations can be seen for all the interactions of these Fe layers, the strongest contribution to the oscillatory behaviour comes from the ferromagnetic NN out-of-plane interactions.  This is to be expected as the variation in the interlayer distance (due to the geometrical relaxation) mostly affect the hybridisation between orbitals centered at adjacent atomic layers. \\*

\subsection{Magnetocrystalline anisotropy energy}
\begin{figure}[ht]
\begin{center}
\includegraphics[scale=0.56,trim=45mm 40mm 20mm 45mm, clip]{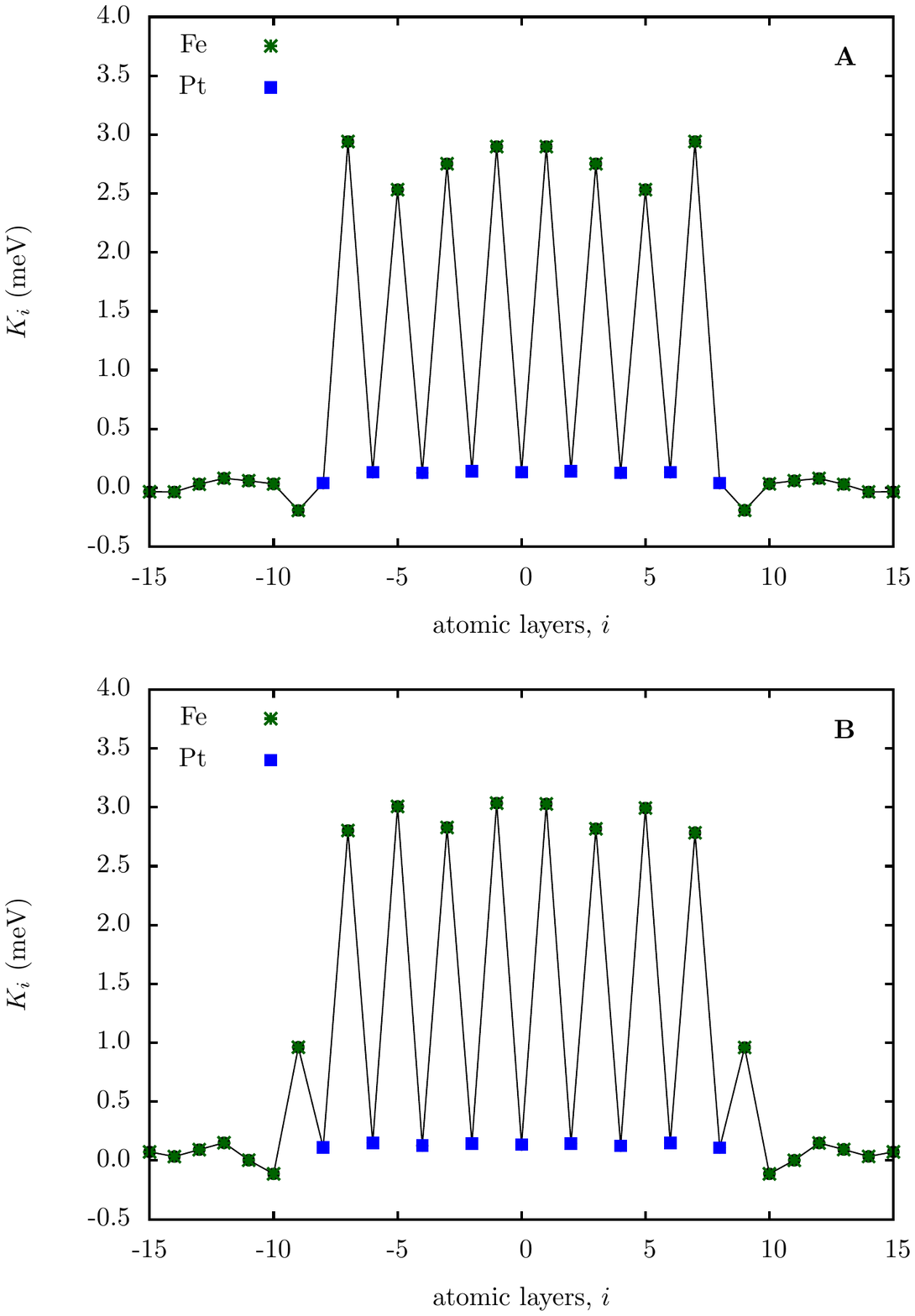}
% \scalebox{0.65}{\input{MAE_UR}}
% \scalebox{0.65}{\input{MAE_CASTEPlargea}}
\end{center}
\vskip -8pt
\caption{Calculated layer-resolved contributions to the MAE, $K_i$, (blue $\blacksquare$:  Pt,  green $\ast$: Fe) for systems {\bf A} and {\bf B}.  Solid lines serve as guides for the eye. }
\label{fig:relMAE}
\end{figure}

Fig.~\ref{fig:relMAE} shows the layer-resolved MAE contributions, $K_i$, see  Eq.~\ref{eq:Kdecomp}, for systems {\bf A} and  {\bf B}.  The MAE contributions of the Fe layers in the FePt slab oscillate between about 2.5 meV and 3 meV. The frequency and the magnitude of these oscillations is different for the two systems, most probably, due to the different boundary conditions.  In system {\bf B}, due to the evidently smaller amplitude of oscillation, $K^{\mathrm{Fe}}_i$ settles more quickly around the bulk value of 3 meV.  In both systems,  $K_i$ of all the Pt layers is very small, $\sim 0.2$ meV. In the Fe parts of the system $K^{\mathrm{Fe}}_i$ quickly approaches (practically) zero, since the MAE of Fe bulk is on the order of $\mu$eV. \\*

\begin{figure}[ht!]
\begin{center}
\includegraphics[scale=0.75,trim=45mm 10mm 20mm 45mm, clip]{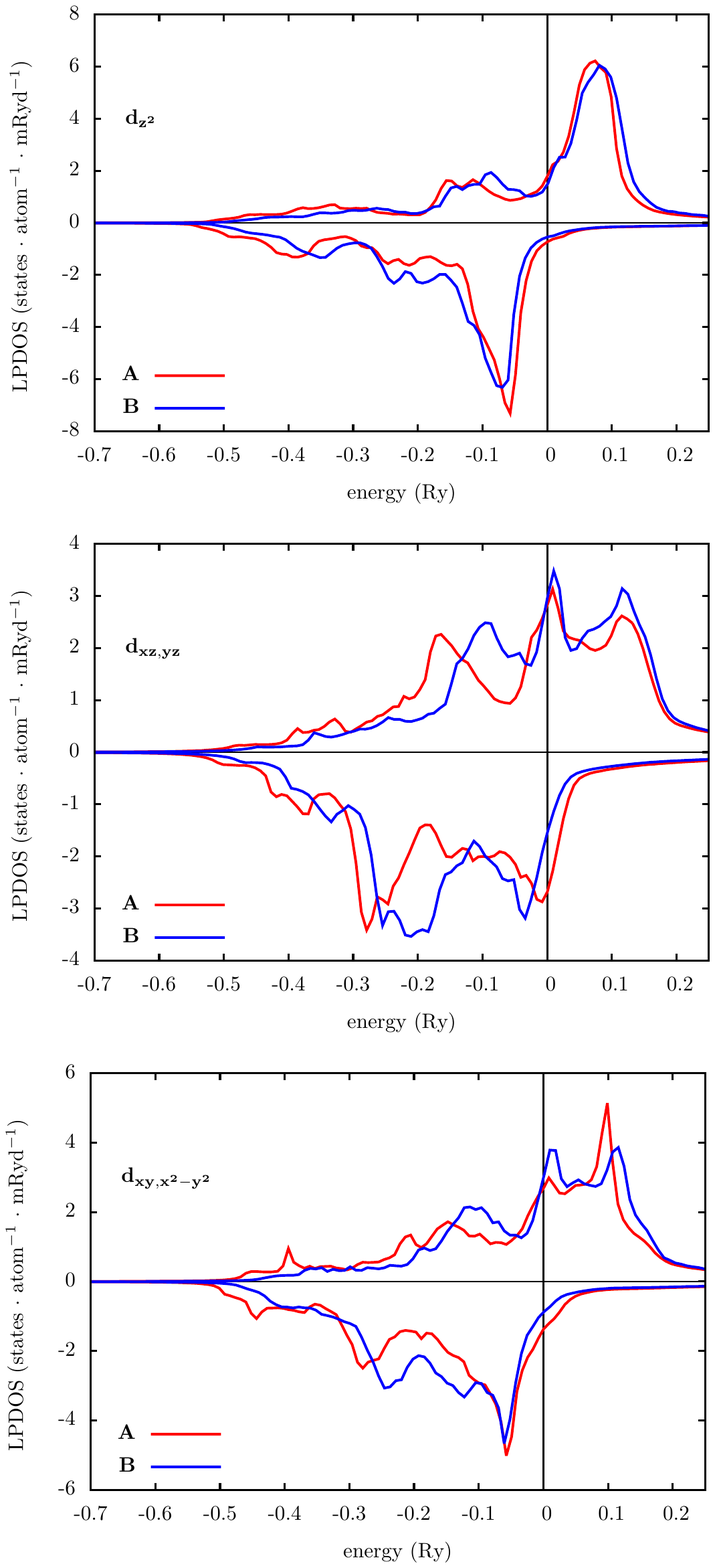}
% \scalebox{0.7}{\input{PDOSd0-6}} \\
% \scalebox{0.7}{\input{PDOSdm1-6}} \\
% \scalebox{0.7}{\input{PDOSdm2-6}}
\caption{Calculated $d$-like spin- and orbital-resolved local partial densities of states (LPDOS) for layer 9. Upper: $d_{z^2}$ ($m=0$), middle: $d_{xz}$ and $d_{yz}$ ($m=\pm 1$), lower: $d_{xy}$ and $d_{x^2-y^2}$ ($m= \pm 2$). Positive/negative values stand for the minority/majority spin-channels. The zero of the energy is shifted to the Fermi energy.}
\label{fig:PDOSdlay9}
\end{center}
\end{figure}

As a remarkable difference between the systems {\bf A} and {\bf B}, in system {\bf B} the Fe layers at the interface (layers $\pm 9$) exhibit a contribution of about 1 meV to the MAE, while in system {\bf A} this contribution is even negative ($\sim -0.25$ meV).  In order to gain an understanding of this difference, we tried applying Bruno's arguments in terms of second order perturbation theory \cite{bruno, ujfalussy-CoAu}.  To this end, we performed self-consistent scalar-relativistic calculations (i.e., calculations in which the spin-orbit coupling is excluded) and calculated the local partial densities of states (LPDOS). In Fig.~\ref{fig:PDOSdlay9} we plotted the $d$-like spin- and orbital-resolved LPDOS at layer 9.  The LPDOS clearly shows a strong spin-polarisation (almost filled majority-spin band) in this layer, which is a necessary condition to apply Bruno's theory.  Apparently, the $d_{z^2}$-LPDOS is nearly insensitive to the geometry relaxation, while upon relaxation a considerable weight of the $d_{xz,yz}$-LPDOS (and, to some extent, also of the $d_{xy,x^2-y^2}$-LPDOS) is shifted  towards the Fermi level in the occupied regime in both spin-channels.  The unoccupied part of the minority-spin channel of these orbital-resolved states is also affected by the geometrical relaxation.  However, in the vicinity of the Fermi level only a small increase in the unoccupied minority-spin LPDOS occurs. (The majority-spin LPDOS clearly decreases at the Fermi level, but it is not relevant in our present theoretical estimation.)  Since the spin-orbit interaction gives rise to couplings between the $d_{xz}$ and $d_{yz}$ states, inducing a perpendicular MAE, as well as to couplings between the $d_{xz,yz}$ and  $d_{z^2}$ states, inducing an in-plane MAE \cite{ujfalussy-CoAu}, it is hardly possible to identify a well-established difference in the specific local contribution to the MAE regarding the two systems. \\*

\begin{figure}[htb]
\begin{center}
\includegraphics[scale=0.56,trim=45mm 40mm 20mm 45mm, clip]{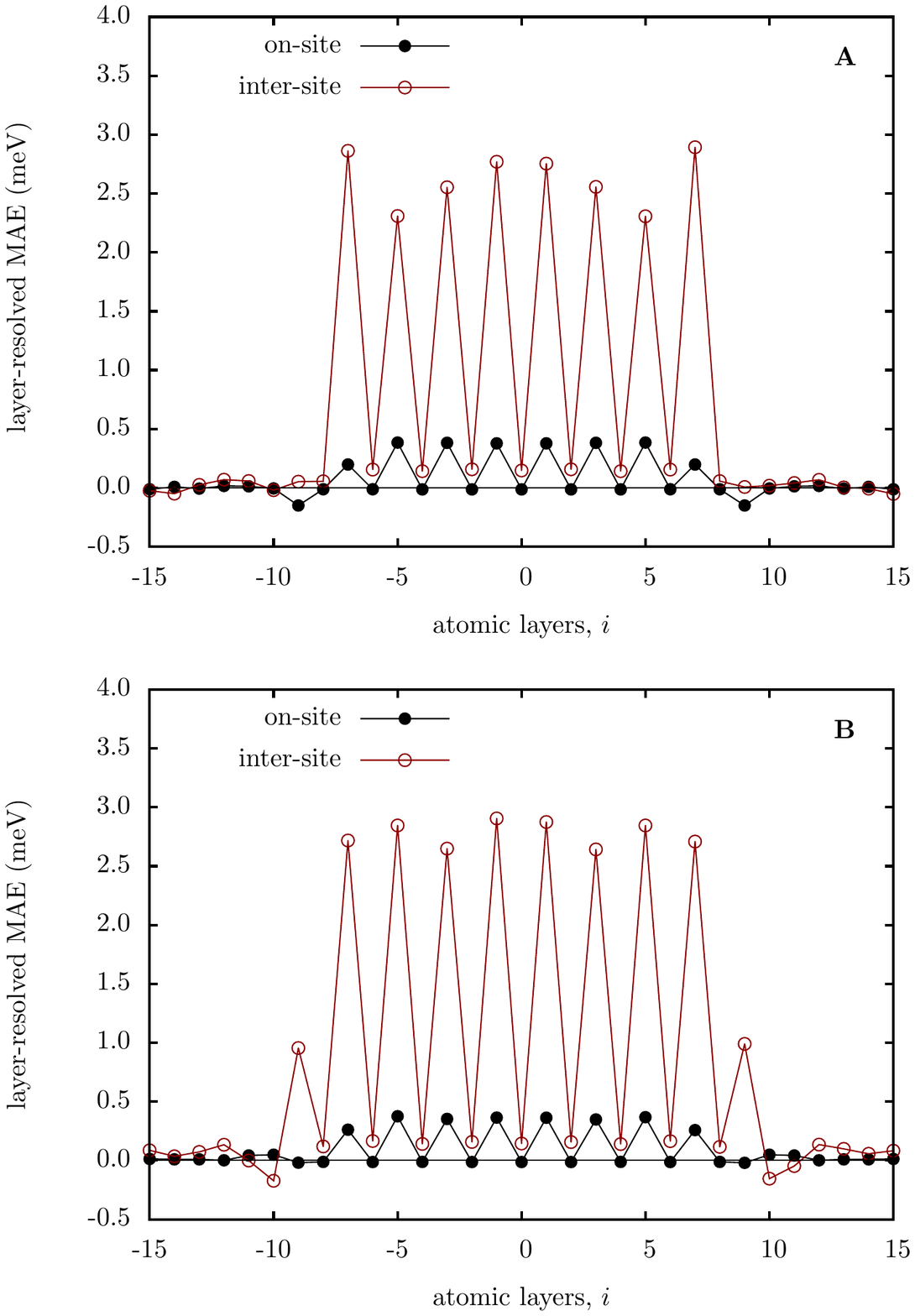}
% \scalebox{0.65}{\input{onsite_twosite}}
% \scalebox{0.65}{\input{OneTwo_largea}}
\end{center}
\caption{Calculated layer-resolved on-site ($\bullet$ ) and inter-site ($\circ$) anisotropies for systems {\bf A} and {\bf B}, see Eq.~(\ref{eq:Ki-spinmodel}).  Solid lines serve as guides for the eye.  \label{fig:onsitetwositeUR}}
\end{figure}

In terms of the localised-spin model as described in Section~\ref{sec:calcs}, the MAE can be cast into on-site and inter-site contributions as per Eq.~(\ref{eq:layerresKi}).  Although the microscopic model to construct an anisotropic spin model differs from the one used in this work, the results of Mryasov \textit{et al.} \cite{mryasov} strongly suggest that the MAE of the FePt systems arise mainly from effective Fe-Fe inter-site interactions, Eq.~(\ref{eq:MAEintersite}), mediated by the spin-orbit coupling on the Pt atoms.  Mapping our results on the MAE to the spin-model, see Eqs.  (\ref{eq:MAEspinmodel}) -- (\ref{eq:Ki-spinmodel}), provides a unique opportunity to check the assertion of Ref. \cite{mryasov}. Using the methods introduced in Ref.~\cite{spinwaves}, we calculated the on-site and inter-site parts of the layer-resolved MAE related to an extended Heisenberg spin model and plotted these contributions in Fig.~\ref{fig:onsitetwositeUR}.\\*

Inspecting Fig.~\ref{fig:onsitetwositeUR} it is indeed obvious that approximately 90 \% of the MAE of FePt is associated with anisotropic inter-site interactions between Fe atoms. The shape of the $K_{i,\mathrm{inter-site}}$ across the atomic layers $i$ coincides reasonably well with that of $K_i$ presented in Fig.~\ref{fig:relMAE}.  The on-site anisotropies are quite stable on the Fe sites within the FePt slab and practically vanish at the Pt sites.  Furthermore, the increase in $K_i$ at the Fe/FePt interface (i.e., in layers $i=\pm 9$) due to relaxation stems mostly from the inter-site anisotropy.  Using the arguments of Mryasov \textit{et al.} \cite{mryasov}, this can be explained in terms of the electron scattering between these Fe atoms and the Pt atoms in the adjacent layer $i = \pm 8$, experiencing thus large spin-orbit coupling. Interestingly, in system {\bf A} this induced anisotropy effect is suppressed and the small negative contribution to the MAE in these layers is of on-site origin.  Remarkably,  Eq.~(\ref{eq:Ki-spinmodel}) is satisfied with a good accuracy if $K_i$ is taken from the direct calculation via the torque method. This lends substantial credit to the use of the tensorial exchange interactions in spin-dynamics simulations. \\*

As a final step in our ab initio calculations, we would like to address the question of how much the MAE of the finite FePt slab is changed with respect to the bulk MAE related to an FePt layer of the same size.  Clearly, this point has a crucial technological impact, namely, whether the perpendicular MAE can be increased by forming an Fe/FePt multilayer sequence.  Regarding  Fig.~\ref{fig:relMAE},  it is obvious that for the chosen width of the FePt layer the effect of a single interface can hardly be separated, since the oscillations of $K_i$ indicate strong interaction between the two Fe/FePt interfaces (quantum interference effects). Thus we define the excess MAE generated by the entire FePt slab as
\begin{equation}
 \Delta K_{\mathrm{FePt-slab}} =  \sum_{ i=-15}^{15} K_i -9 K_{\mathrm{FePt}} \; ,
 \label{eq:MAEslab}
\end{equation}
where in the sum we also include layers from the Fe part of the system. Note that the MAE of Fe bulk (on the order of $\mu$eV/atom) is neglected and $K_{\mathrm{FePt}}$ is the MAE per formula unit (f.u.) for bulk FePt. We calculated $K_{\mathrm{FePt}}=3.37$~meV/f.u., which, although high in comparison to experiment, is in good agreement with other theoretical results based on the LSDA or the LSDA+U approach \cite{lyubina,shick}. \\*

When evaluating $\Delta K_{\mathrm{FePt-slab}}$ we needed to consider that for the systems {\bf A} and {\bf B} the Fermi level of bulk Fe is used instead of the Fermi level of bulk FePt, which slightly affects the value of $K_{\mathrm{FePt}}$ calculated within the SKKR-ASA approach. In order to calibrate $K_{\mathrm{FePt}}$ we used system {\bf  C} (an unrelaxed Fe/FePt/Fe trilayer immersed in FePt bulk, see Section~\ref{sec:calcs}). Indeed, in the FePt slab of system {\bf C} we obtained a very similar shape of $K_i$ across the atomic layers as for system {\bf A}, while for the innermost FePt layers the bulk MAE, $K_{\mathrm{FePt}}$, was retained to within less than 1~\% numerical accuracy. Since the corresponding MAE contributions in system {\bf A} are by 0.33 meV/f.u. smaller, in Eq.~(\ref{eq:MAEslab}) we used a corrected value of 3.04 meV/f.u for $K_{\mathrm{FePt}}$.  \\*

For system {\bf A}, we obtain a reduction in the total MAE, $\Delta K_{\mathrm{FePt-slab}}^{\mathrm{\bf A}} \approx -4.2$ meV.  This reduction stems primarily from the interfacial layers $i = \pm 9$, see Fig.~\ref{fig:relMAE}.  From this figure it is obvious that the contribution of one Fe layer to the MAE of the FePt slab is 'missing', since this Fe layer becomes much rather an interfacial Fe layer. Upon relaxation, i.e. in system {\bf B}, the interfacial Fe layers have remarkably enhanced contributions to the MAE, see Fig.~\ref{fig:relMAE}, as these Fe layers seem to belong rather to the FePt slab. Moreover, in this case the Fe layers in the FePt slab have contributions to the MAE closer to that in bulk FePt as compared to system {\bf A}. Consequently, for system {\bf B} the MAE of bulk FePt is almost entirely retained, $\Delta K_{\mathrm{FePt-slab}}^{\mathrm{\bf B}} \approx -0.4$ meV. Comparing this value to the total MAE of the FePt slab immersed in Fe, 26.9~meV, we conclude that
the MAE of a realistic (Fe$_m$/(FePt)$_k$)$_n$ ($m \gtrsim 10$, $k \gtrsim 9$) multilayer sequence is approximately equal to the MAE of $n \cdot k$ FePt bulk layers.

\section{Implications for mesoscopic spin structures}
The detailed mapping of the ab-initio information onto a spin model, which will allow, for example, calculations of the temperature dependence of the MAE values, is beyond the scope of the current work. Here we give a simple illustration of the implications of the ab initio results and their effect on magnetic spin structures and indeed the exchange spring phenomenon. Specifically, we consider the effect of an abrupt change in magnetic properties, especially the MAE, which is known to give rise to pinning of a domain wall at the interface. Kronm\"{u}ller and Goll \cite{kron1} developed a micromagnetic model of magnetization reversal in a material consisting of two coupled phases with different magnetic properties. It was found that a domain wall (DW) could be pinned at the interface between the different layers, the pinning being overcome by a critical field
\begin{equation}
H_c= \frac{2K^{II}}{M_s^{II}}\frac{1-\epsilon_K\epsilon_A}
{(1+\sqrt{\epsilon_M\epsilon_A})^2} \: ,
\label{eq:kron}
\end{equation}
where the superscript $II$ refers to the properties of the hard phase and $\epsilon_K,\epsilon_A,\epsilon_M$ refer to the ratio of the anisotropy constant, the micromagnetic exchange constant and saturation magnetization respectively in the soft and hard phases. The coercivity of the hard phase is invariably reduced by all combinations of material parameters.

However, Eq.~(\ref{eq:kron}) was derived under the assumption of bulk exchange coupling across the interface, and it has been shown by Guslienko {\it et al.} \cite{gus} that the exchange spring effect is strongly dependent on the degree of coupling at the interface. In Ref.~\cite{gus} the interface coupling was taken as a variable, but the ab-initio calculations presented here allow to study the exchange spring phenomenon with no fitting parameters. We use a simple spin-chain model as in Ref.~\cite{gus}, treating the low-exchange layer as an interface providing a weakened exchange between the FePt and Fe layers. Within each layer we can write down the following spin Hamiltonian
 \begin{equation}
  \mathcal{H} = -J \sum_{i,j (nn)} \vec{S}_i \cdot \vec{S}_j + \sum_i K (S_i^z)^2 - \sum_i \mu \vec{H}\cdot  \vec{S}_i \: ,
 \label{eq:layer_H}
 \end{equation}
 where $J$ is the intralayer nearest-neighbor exchange coupling,  $\vec{S}_i$ the unit vector representing the spin direction, $K$ the anisotropy constant, $\mu$ the atomic spin in the given layer and $\vec{H}$ the applied field.\\*
 
We allow for reduced exchange coupling at the interfaces by writing the exchange energy between interface spins as
 \begin{equation}
  \mathcal{H}_{int} =- J_{int} \sum_{i,j} \vec{S}_i \cdot \vec{S}_j,
  \label{eq:coupl}
 \end{equation}
 where the spins $i,j$ are in separate layers.   Similar to the case of IrMn$_3$/Co $(111)$ system in Ref.~\cite{IrMnCo}, our calculations show sizeable DM interactions near the interface.  Due to the C4v symmetry, however, the DM energy cancels for the N\'eel walls to be investigated in the following.   For that reason, we neglected the DM interactions in our model of Eq.~(\ref{eq:coupl}). The equilibrium state of the spin system is determined by integrating the Landau-Lifshitz equation, without the precession term
\begin{equation}
\frac{d\vec{S}_i}{dt}=-\alpha\,\vec{S}_i\times\left(\vec{S}_i\times\vec{H}_i\right) \: ,
\end{equation}
with $\vec{H}_i$ being the effective field acting on spin $i$.
\begin{figure}[htp!]
\centering
\includegraphics[scale=0.56,trim=45mm 145mm 20mm 45mm, clip]{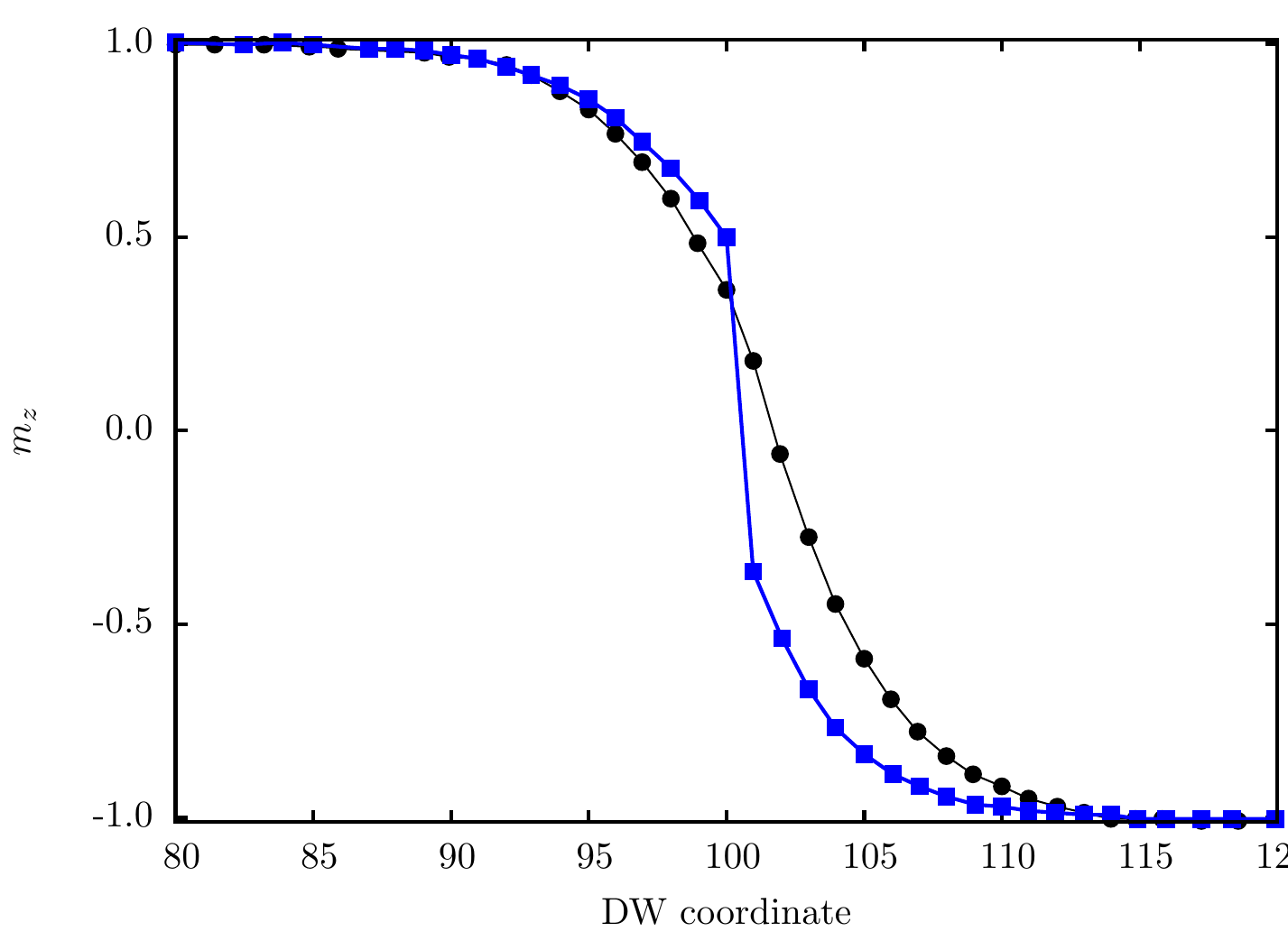}
% \scalebox{0.65}{\input{DWfig}}
\caption{Calculated Domain Wall structures prior to magnetization reversal for fully exchange coupled layers (black $\bullet$) and layers coupled with ab-initio exchange parameters for the relaxed system {\bf B} (blue $\blacksquare$). The relaxed system shows a discontinuous DW structure due to the weak interlayer exchange coupling. \label{fig:DW_profiles}}
\end{figure}

Fig.~\ref{fig:DW_profiles} shows calculated Domain Wall (DW) structures prior to magnetization reversal for fully exchange coupled layers and layers coupled with ab-initio exchange parameter for the relaxed system. In the latter case the interlayer exchange is approximately 20\% of the bulk exchange of FePt. The associated coercivities are 5.9~T and 7.24~T respectively, reduced from the bulk coercivity of 14~T due to the exchange spring effect. For the bulk interlayer exchange case the DW width in the Fe layer is considerably smaller than the usual expectation due to the presence of the large applied field. Nonetheless the DW is continuous across the interface, in contrast to the case of the reduced exchange of the relaxed microstructure. The effect of the reduced exchange on the coercivity is relatively weak, consistent with the results given in Ref.~\cite{gus}. However, even relatively small changes can be significant for the design and operation of practical recording media.

\section{Conclusion}
We have presented first principles calculations of the exchange interactions and the magnetocrystalline anisotropy energy (MAE) in an Fe/FePt/Fe sandwich system. In particular, we investigated how the geometrical relaxation influences the calculated magnetic properties of the system.  In accordance with previous work on an unrelaxed Fe/FePt/Fe system \cite{sabiryanov}, we found a dramatic reduction in the exchange coupling between the Fe layers at the Fe/FePt (soft/hard) interface.  Moreover, in the relaxed system, these layers add a remarkable positive contribution to the MAE. From the tensorial exchange interactions evaluated by means of the relativistic torque method \cite{spinwaves} we have shown that the MAE of the FePt slab and the interface MAE are dominated by anisotropic inter-site exchange interactions. Moreover, our calculations indicate that the formation of an Fe/FePt layer sequence reduces the perpendicular MAE.  In the case of a relaxed geometry, which we consider to be relevant to a multilayer system, this reduction is slight, on the order of $\sim 0.4$ meV per FePt slab. \\*

We also show that the ab-initio parameters will have a bearing on the mesoscopic spin structures predicted by atomistic and micromagnetic models. Specifically, the reduced exchange coupling between FePt and Fe layers gives rise to a discontinuous spin structure across the FePt/Fe interface. Although the exchange spring effect still gives rise to a large coercivity reduction it is likely that the discontinuous spin structure could affect the magnetization dynamics. The reduced exchange could also affect the temperature dependence of the interface MAE values, which could also have a significant bearing on the magnetic properties. Both factors require the development of a detailed mapping onto an atomistic spin model. Further, each material combination in a magnetic nanostructure will have interface properties which may differ significantly from the bulk, making further ab-initio studies of interface properties important in terms of the understanding of the underlying physics of static and dynamic magnetic properties.

\section{Acknowledgments}
Financial support was provided by the Hungarian National Research Foundation (under contracts OTKA 77771 and 84078) and by the New Sz\'echenyi Plan of Hungary (Project ID: T\'AMOP-4.2.2.B-10/1--2010-0009).  Support of the EU under FP7 contract NMP3-SL-2012-281043 FEMTOSPIN is gratefully acknowledged.  CJA is grateful to EPSRC for the provision of a research studentship.\\*

\end{document}